%% file: sample-sigconf.tex
\documentclass[sigconf, authorversion, nonacm]{acmart}

\AtBeginDocument{%
  \providecommand\BibTeX{{%
    \normalfont B\kern-0.5em{\scshape i\kern-0.25em b}\kern-0.8em\TeX}}}

\setcopyright{acmcopyright}
\copyrightyear{2023}
\acmYear{2023}
\acmDOI{XXXXXXX.XXXXXXX}

\acmPrice{}
\acmISBN{978-1-4503-XXXX-X/18/06}

\usepackage{xspace}
\usepackage{cleveref}
\usepackage{caption}
\usepackage{subcaption}
\usepackage{assets/card-template}
\usepackage{preprint_header}

\newcommand{\cardsName}{AI Usage Cards\xspace}

\begin{document}
\title{\cardsName: Responsibly Reporting AI-generated Content}

\author{Jan Philip Wahle}
\affiliation{%
  \institution{University of Göttingen}
  \city{Göttingen}
  \country{Germany}
  \postcode{37073}
}
\email{wahle@uni-goettingen.de}

\author{Terry Ruas}
\affiliation{%
  \institution{University of Göttingen}
  \city{Göttingen}
  \country{Germany}
  \postcode{37073}
}
\email{ruas@uni-goettingen.de}

\author{Saif M. Mohammad}
\affiliation{%
  \institution{National Research Council Canada}
  \city{Ottawa}
  \country{Canada}
  \postcode{XXXXXX}
}
\email{saif.mohammad@nrc-cnrc.gc.ca}

\author{Norman Meuschke}
\affiliation{%
  \institution{University of Göttingen}
  \city{Göttingen}
  \country{Germany}
  \postcode{37073}
}
\email{meuschke@uni-goettingen.de}

\author{Bela Gipp}
\orcid{0000-0001-6522-3019}
\affiliation{%
  \institution{University of Göttingen}
  \city{Göttingen}
  \country{Germany}
  \postcode{37073}
}
\email{gipp@uni-goettingen.de}

\renewcommand{\shortauthors}{Wahle et al.}

\begin{abstract}
Given AI systems like ChatGPT can generate content that is indistinguishable from human-made work, the responsible use of this technology is a growing concern. Although understanding the benefits and harms of using AI systems requires more time, their rapid and indiscriminate adoption in practice is a reality. Currently, we lack a common framework and language to define and report the responsible use of AI for content generation. Prior work proposed guidelines for using AI in specific scenarios (e.g., robotics or medicine) which are not transferable to conducting and reporting scientific research. 
Our work makes two contributions: First, we propose a three-dimensional model consisting of transparency, integrity, and accountability to \textit{define} the responsible use of AI. 
Second, we introduce ``AI Usage Cards'', a standardized way to \textit{report} the use of AI in scientific research. 
Our model and cards allow users to reflect on key principles of responsible AI usage. They also help the research community trace, compare, and question various forms of AI usage and support the development of accepted community norms.
The proposed framework and reporting system aims to promote the ethical and responsible use of AI in scientific research and provide a standardized approach for reporting AI usage across different research fields.
We also provide a free service to easily generate AI Usage Cards for scientific work via a questionnaire and export them in various machine-readable formats for inclusion in different work products at \url{https://ai-cards.org}.
\end{abstract}



\begin{teaserfigure}
  \vspace*{-3mm}
  \centering
  \begin{subfigure}[c]{0.48\textwidth}
     \centering
     \includegraphics[width=\textwidth]{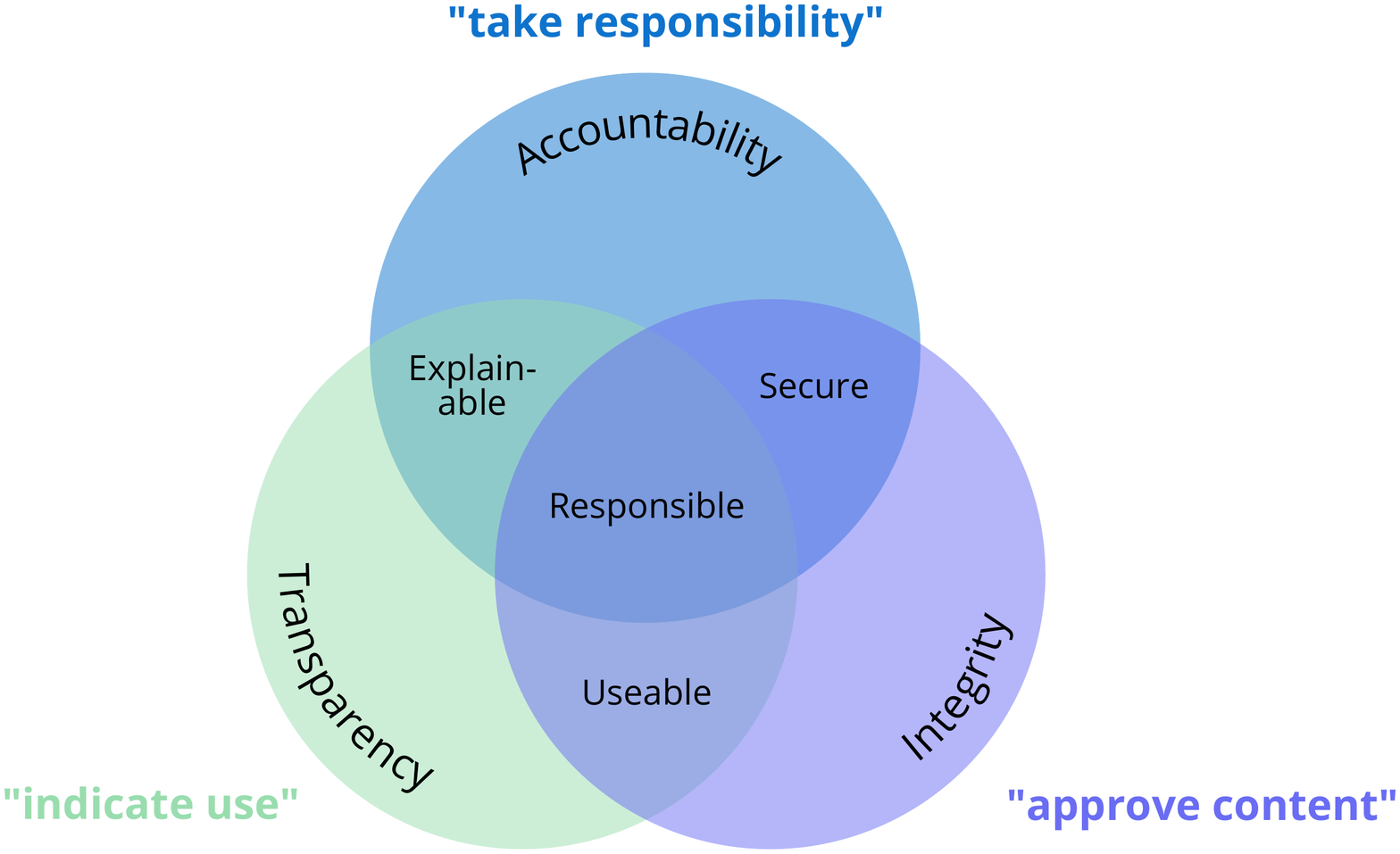}
 \end{subfigure}
 \hfill
 \begin{subfigure}[c]{0.48\textwidth}
     \centering
     \includegraphics[width=\textwidth]{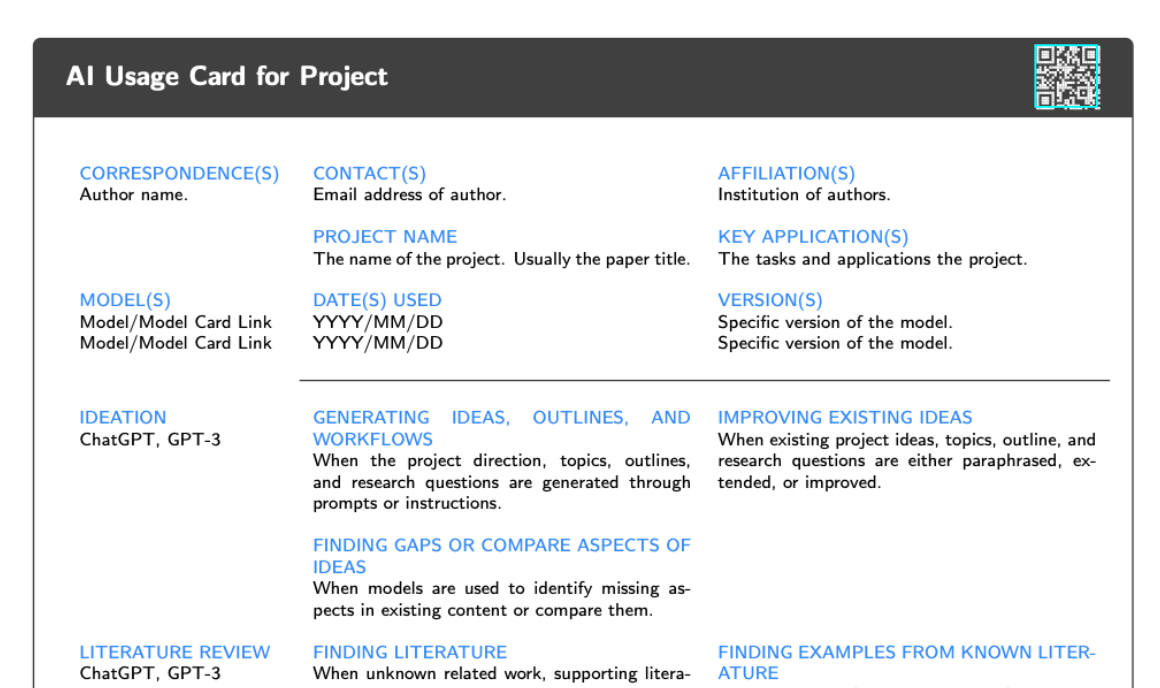}
 \end{subfigure}
 \hfill    
  \vspace*{-6mm}
\caption{A three-dimensional model of responsible AI usage (left) and the header of the \cardsName template (right).}
  \label{fig:teaser}
  \vspace*{3mm}
\end{teaserfigure}

\maketitle

\thispagestyle{preprintbox}

\section{Introduction}

The rapid development of AI for natural language processing has led to concerns about using AI-generated content in scientific papers. While AI systems like ChatGPT \cite{chatgpt} can generate texts, ideas, and source code, some outputs may be false or fail to attribute the appropriate sources. As a result, it remains unclear whether such models should be used for content generation.

Despite the novelty of AI for content generation, researchers are already using it to support software development and academic writing. However, the rules and norms to govern this new paradigm are yet to be determined. Conferences and journals have started to define rules for using language models to generate content. The largest conference in computational linguistics (ACL) differentiates between AI-based support tools in its official language model policy \cite{aclpolicy} and calls for \textit{transparency}, i.e., acknowledging the use of AI when generating content. Conversely, one of the most significant machine learning conferences (ICML) has prohibited using all language models for submissions in 2023 \cite{icmlpolicy}. Although these attempts are not a final and unanimous agreement on using AI-based assistants, they do signal a response to the sudden increase in AI content generation.

While the rules and norms to govern AI-generated content are still emerging, it is clear that transparency is an essential element of responsible use. By acknowledging the use of AI, users can ensure that the generated content is reliable and appropriate attribution is given to the appropriate sources. However, the responsible use of AI for content generation is context-dependent and complex, and caution must be exercised in fields where high-stakes outcomes are involved, such as in the medical and legal fields. Ultimately, while using AI for content generation is likely here to stay, we must find ways to use it responsibly and transparently to avoid adverse outcomes.

History has shown that carefully scoped legalizations can be more effective than blanket prohibitions in many scenarios.
Similar to current discussions on social issues such as recreational drug use, a formal framework for acceptable use creates transparency, helps to control the problem, and decriminalizes usage. 
However, we lack a common language and concrete principles to describe what is needed to ``responsibly legalize'' AI usage.
Although frameworks exist to document key characteristics and ethical considerations associated with models \cite{mitchell2019}, datasets \cite{pushkarna2022}, evaluations \cite{hupkes2023}, and AI tasks \cite{mohammad-2022-ethics}, in the form of cards and sheets, no standardized framework exists to report the use of AI-generated content.

In this work, we introduce the three-dimensional model shown in \Cref{fig:teaser} to decompose the dimensions of responsible AI usage for content generation.
We use the term ``AI Usage'' to refer to all applications of AI-based assistants in scientific work, e.g., to inspire, generate, revise, and compare content.
In addition to \textbf{transparency} (``Where and how was AI used to produce content?''), our model incorporates \textbf{integrity} (``Have humans approved the content produced by AI systems?'') and \textbf{accountability} (``Who is responsible for the use/dissemination of content produced by AI systems?'').

We also propose \textit{\cardsName} as a standard for acknowledging AI support in the scientific process---an important scenario among the many in which content generation can be used.
We crafted \textit{\cardsName} with the phases of the scientific process in mind, e.g., hypothesis definition, literature review, and manuscript writing. However,  we designed the framework flexible enough to report AI usage across domains.
Our cards seek to make the acknowledgment of AI usage the status quo.
In summary, this paper makes the following two contributions.

\begin{enumerate}
    \item We propose a model for the \textit{\textbf{responsible}} use of AI content generation in science based on three dimensions: \textbf{\textit{transparency}}, \textbf{\textit{integrity}}, and \textbf{\textit{accountability}}.
    \item We introduce \cardsName, as a standardized tool to report using AI in scientific works. \cardsName can be generated using an online questionnaire, which we provide as a free service and exports in machine-readable formats that can be incorporated directly into any scientific work.
\end{enumerate}
\noindent Our proposed model and method of documenting AI usage can serve as a valuable tool for users to engage in the responsible use of content-generating systems. However, it is important to recognize that potential scenarios for using AI in scientific work can vary significantly, ranging from low-stakes tasks, such as correcting typos in a social media post, to high-stakes scenarios that have the potential to cause harm, such as misleading diagnoses or unjustified sentencing, especially in the medical and legal fields.

In these high-stakes scenarios, AI should be used with extreme caution, and careful consideration should be given to the potential consequences of AI-generated content. There may be applications that we do not want to use AI for at all. In addition, the target audience should also be considered, whether experts or laypeople, as they perceive information as more or less critical. We want to emphasize that simply legalizing AI usage is not the solution; rather, we must find ways to define where and how we should use it and make its usage responsible for avoiding negative consequences.

While people will undoubtedly continue to use AI for content generation, we must take extra care to ensure its responsible use. If used transparently and responsibly, we believe AI can do more good than harm.

\section{Related Work}

For some time now, studies have shown that humans are increasingly unable to detect AI-generated content \cite{brown2020language, Wahle2022d, Guo2023}.
Efforts exist to flag generated text automatically, for example, through zero-shot detection methods \cite{Mitchell2023DetectGPT} or watermarks introduced in generated texts \cite{Kirchenbauer2023LMWatermark}.
However, detection techniques alone do not solve important aspects of AI usage, e.g., integrity and accountability.

Concerns regarding the use of AI are also present in the health domain for medical diagnosis \cite{NeriCMB20}, clinical trials \cite{CruzRiveraLCD20}, and health care research \cite{IbrahimLD21}.
When providing diagnosis, recommending treatment, and screening patients, AI models (such as ChatGPT) must have their predictive capabilities strictly constrained to avoid serious harm, as these models contain biases and unverified information \cite{LiuCMC20}.
Moreover, clear accountability for actions taken based on recommendations from AI is required, i.e., ``Who is responsible for the diagnosis and treatment?''.
Although relevant, the contributions of \citet{NeriCMB20,CruzRiveraLCD20,LiuCMC20,IbrahimLD21} are either not transferable to other domains, lack a common agreement, or miss artifacts relevant to our use case in their reports, e.g., tracking which models are used and transparency on modified textual content.
Nonetheless, these works mostly agree on the need for better reporting guidelines and more responsible AI.

\cardsName is in line with contributions such as Data Cards \cite{pushkarna2022}, Model Cards \cite{mitchell2019}, and Evaluation Cards \cite{hupkes2023}, which complement our work.
Data Cards highlight the most important facts about machine learning datasets.
The goal of a Data Card is to inform those involved (e.g., researchers) about the characteristics of manipulated datasets during their lifecycle in specific projects (e.g., content, number of records, supporting funding agency).
Model Cards \cite{mitchell2019} document machine learning models---mainly from computer vision and natural language processing---to provide the necessary details for reproducing the results, e.g., training data, metrics, and model parameters. 
In addition to the technical details on a specific model, \citet{mitchell2019} advocate for transparency of the performance of considered models.
Evaluation Cards \cite{hupkes2023} allow researchers to document their experiments and standardize evaluations, thereby enabling the monitoring of trends in NLP research.

\section{A Three-Dimensional Model} \label{sec:model}
To responsibly use the outputs of AI, we propose categorizing its usage in a three-dimensional model. 
The model includes pillars to help acknowledge the usage of AI models (\textit{transparency}); to approve the generated content (\textit{integrity}); and clarify who is responsible for the content generated by the AI model (\textit{accountability}).

While specific frameworks for the use of AI exist in some areas (e.g., public decision-making \cite{deFineLichtd20}, robotics \cite{WachterMF17}), they are still limited as they do not specify to which aspects of the scientific work AI systems contributed.
Our model is general enough to be applied to many different research domains and still categorizes the main components of any scientific work.
In the following, we detail each dimension of our conceptual model and explain how they interact.\\[3pt]
\noindent \textbf{Transparency.} The first dimension of responsible practice is the foundation on which all others build.
Transparency refers to acknowledging the use of AI models in one's work.
Throughout the project, one should document every instance of using AI models, with the intent to disclose the information in a suitably summarized form eventually.
Similar practices are established for acknowledging a funding agency for supporting a project, individuals for giving feedback, an author affiliation, or a possible conflict of interest.
The primary goal of disclosing relevant information about AI usage during scientific work is to establish traceability of the scientific process steps to which AI contributed  and help authors reflect on which parts of their work have originated from AI.\\[3pt]
\noindent 
\textbf{Integrity.} The second dimension pertains to the correctness, truthfulness, fairness, safety, and appropriateness of the content generated by AI systems.
AI uses learnable parameters to model massive amounts of data from the internet.
Thus, users of this technology should reflect on such content and assert whether it propagates biases, factual incorrectness, other's ideas as own ideas, unethical statements, prejudice, personally identifiable information, or false claims towards individuals in their output.
This verification of fairness principles has to be carried out through human assessment.
For smaller initiatives, we recommend obtaining input from a diverse set of people, with representation from relevant groups of people most impacted by system bias. 
For larger projects, we recommend adopting strategies from participatory design.
Participatory design in research and systems development centers the people, especially marginalized and disadvantaged communities, such that they are not merely passive subjects but have the agency to shape the assessment process \cite{spinuzzi2005methodology,humphries2020arguments,noel2016promoting, oliver1997emancipatory}. 
All aspects of integrity assessment must be documented for eventual reporting (as appropriate). \\[3pt]
\noindent \textbf{Accountability.} The last dimension addresses how researchers use, publish, and disseminate content produced with the help of AI models.
With clear accountability, those affected by serious harm can ask authors to fix overlooked integrity failures (e.g., propagating hate speech content), especially for those already marginalized.
On one side, accountability means ensuring that generated content adheres to integrity proactively before releasing AI-generated content.
On the other side, it means being open to feedback and willing to make changes as necessary, and being open about the model’s limitations and potential risks associated with its use.
Many criticize that AI systems such as ChatGPT are largely black boxes that need more transparency, not least because there has been no official research paper associated with its release. 
However, transparency is often confused with accountability.
When humans perform inexplainable decisions, we can hold them accountable, but if a search engine provides racist results, we cannot.
Without accountability, knowing what was manipulated without accountability can produce harm to individuals and does not lead to liability.
Therefore, responsible AI use requires documenting corresponding individuals for the project who can be contacted about potential issues.\\[3pt]
\noindent \textbf{Responsibility.}
Ideally, all three dimensions, i.e., transparency, integrity, and accountability, must be fulfilled simultaneously for using AI models responsibly.
Analogous to a three-legged chair, each leg/dimension is crucial for using AI in the scientific process.
By acknowledging the use of models and holding authors accountable, the output can be explained, but it may still contain inaccuracies and biases since no verification steps have been taken. It is the responsibility of both the AI user and creator to provide sources for the ideas presented, allowing readers to verify them.
If we acknowledge the usage of AI and approve its content, we achieve usability but can encounter a lack of accountability if problems occur, e.g., plagiarized content or misleading information.
The content is useable ``at your own risk'' due to a lack of accountability.
Finally, when accountability and integrity are verified, the output is secure because it has been affirmed. Even if human flaws exist, they can be accounted for and adjusted if necessary.
Only when we acknowledge the use of the model, mitigate its inappropriate outputs, and assign someone to respond in case of problems or questions can we mitigate major issues and make AI usage responsible.

\section{\cardsName}
To facilitate the documentation and reporting needs for the responsible use of AI-generated content, we propose \cardsName as a standardized practice to incorporate the principles of transparency, integrity, and accountability.

The \cardsName presented in this paper are motivated towards using AI in scientific works; however, these cards can easily be adapted to other domains.
\cardsName for scientific works follow the scientific method and reflect common headings of scientific manuscripts.
We plan to offer different card templates for different disciplines---from computer science to zoology---and stakeholders.
Companies, funding agencies, or institutions could define individual questions with varying levels of detail depending on what they deem necessary for ensuring responsible AI usage in their scenario.

\cardsName are available in multiple machine-readable formats, such as \LaTeX, XML, JSON, and CSV, to enable easy inclusion in different work products (cf. \Cref{fig:teaser}) and support the automated analysis of AI usage across domains and content types in the future. \cardsName can be redistributed for non-commercial purposes according to the CC BY-NC 4.0 license\footnote{\url{https://creativecommons.org/licenses/by-nc/4.0/}}. A free service to create \cardsName\footnote{The AI card reporting usage of AI for this paper can be found on the project webpage.} is available at

\begin{center} \textbf{\url{https://ai-cards.org}} \end{center}

\subsection{Structure}
\cardsName are divided into six major blocks: \textit{project details}, \textit{ideation and review}, \textit{methodology and experiments}, \textit{writing and presentation}, \textit{code and data}, and \textit{ethics}.
Our card allows researchers and practitioners to report AI usage during all phases of scientific work, from theoretical ideation to practical use.
This includes even formulating ideas, for example, in the form of questions and acquiring knowledge during a literature review.
Researchers can prompt AI to generate hypotheses, support the development of methods, and suggest newly designed experiments.
For measurements, AI can suggest data sources and process data using generated code.
Finally, when communicating results, AI can support the writing of papers or other presentation artifacts such as figures and tables.
In the following, we summarize each of these blocks:

\begin{itemize}
    \item \textbf{Project details}: Meta-information about the authors, models, and project, including their names, versions, key applications, and affiliations.
\item \textbf{Ideation and review}: Support during theoretical preparation of the project through idea formulation (e.g., outline) and examining related work (e.g., comparing literature, finding analogies).
\item \textbf{Methodology and experiments}: Design, comparison, and generation of processes for methodological tasks during the project, optionally resulting in experiments (e.g., proposing a new process).
\item \textbf{Writing and presentation}: Generation and paraphrasing text used in scientific reports or papers. The improvement of content, figures, tables, and any other elements are also included.
\item \textbf{Code and data}: Manipulation, generation, refactoring, optimization, and analysis of code and data.
\item \textbf{Ethics}: Implications of using AI and steps to mitigate their possible errors and harms.
\end{itemize}
\noindent For more details on each category see \Cref{appendix:cardsample}.
Each block contains six to seven subcategories. In each subcategory, one or more of the following classifications can be assigned.
\begin{itemize}
    \item  \textbf{New:} Content generated based on prompts without a significant portion of prior ideas and thoughts.\\
\textit{Example Prompt:} ``Generate some ideas on how to approach the problem of memorization for large language models.''
\item \textbf{Revise:} Content generated based on previous content that has a significant portion of own ideas and thoughts.\\
\textit{Example Prompt:} ``Rephrase the following paragraph so that it uses academic voice, is concise and short, and adds an example of a meta-analysis.'' 
\item \textbf{Compare:} Content generated by providing two or more pieces of content (own or others).\\
\textit{Example Prompt:} ``Compare my definition of plagiarism to the following [...]''.
\end{itemize}

We devise a questionnaire to generate \cardsName (see \Cref{appendix:questionnaire} for more details).
The questions can be answered through a free online form that automatically generates a card to be incorporated in any scientific report as shown in Card \ref{card:template}.
The AI card that reports usage for our paper can be found on the project webpage.

\section{Practical Considerations}

\noindent \textit{Q1. Who should create \cardsName?}\\[2pt]
\noindent A. All individuals using any AI system to assist in work are eligible to use \cardsName. The use is intended for a broader audience, from students reporting their assignments at school to researchers submitting research articles and funding proposals. \\

\noindent \textit{Q2. When should one create \cardsName?}\\[2pt]
\noindent A. Broadly, \cardsName should be created when using AI systems to support content-related activities during any project (e.g., scientific paper, blog post). This means documenting any activity in which AI has been used as a support tool and categorizing the support according to the card.
\\

\noindent Q3. What are the benefits of \cardsName?\\[2pt]
\noindent A. The benefits of \cardsName include providing transparency, accountability, and integrity in AI usage by encouraging authors to reflect on their AI usage, giving individuals tools to acknowledge AI usage, addressing AI biases for minorities, publishing machine-readable reports for analysis, and contributing to responsible AI usage becoming the status quo. \\

\noindent \textit{Q4. What should the cards not be used for?}\\[2pt]
\noindent A. \cardsName should not be used to justify unethical or inappropriate content dissemination. This card should force users to reflect on the generated content and its value to society.\\

\noindent \textit{Q5. Should \cardsName be updated?}\\[2pt]
\noindent A. Yes. As technologies change and different domains have varying requirements, the cards must be revised periodically to ensure their relevance and mitigation of outdated practices. For example, the concerns raised by the use of large language models today might not be relevant ten years from now. \\

\noindent \textit{Q6. Should we have specific \cardsName?}\\[2pt]
\noindent A. \cardsName should serve as a framework to incorporate different questions for different needs.
Given its machine-readable format, \cardsName will allow for data-driven transparency on what was considered important for responsible AI usage by different groups of people throughout time. There might be different cards for different scientific works (e.g., dissertation---more details; short paper---fewer details). Our card might also find applications in other areas, such as philosophy or art, with varying requirements.\\ 

\noindent \textit{Q7. How can we incentivize the creation of \cardsName?}\\[2pt]
\noindent A. Further simplifying the creation of \cardsName, e.g., by offering use-case-specific card formats, can encourage widespread adoption. Moreover, requiring a report on AI usage for any scientific submission, similar to a statutory declaration in graduation theses, can incentivize authors to internalize responsible reporting practices. \cardsName allow creating such reports efficiently in a standardized and machine-processable fashion.  \\

\noindent \textit{Q8. What role should \cardsName play in the review process of conferences and journals?}\\[2pt]
\noindent A. They should act as documentation to help conference and journal organizers to understand how much support AI systems offer to their users during their work. However, an acceptable threshold for such assistance should be clarified by the organizers. \\

\noindent \textit{Q9. Won't the reporting of \cardsName slow the research work?}\\[2pt]
\noindent A. No. \cardsName can be generated in less than five minutes using a free, interactive, and easy-to-use questionnaire. They will help individuals reflect on how they use AI systems in their work. The responsible use of AI and its report is in the interest of all parties involved \cite{mohammad-2022-ethics}.

\section{Discussion}
In this paper, we proposed \cardsName to facilitate the reporting of AI use. 
We provided a free service to produce cards through a fast and interactive questionnaire.
We showed how AI use could be reported in a standardized way for all steps of scientific work.
Beyond scientific projects, our card serves as a framework to report the usage of AI models across domains.

\cardsName allow to monitor AI usage and help policymakers to evaluate their decisions.
Our card can be exported in machine-readable formats (e.g., XML) for further analysis.
Compared to other efforts for reporting AI usage \cite{NeriCMB20,CruzRiveraLCD20,LiuCMC20,IbrahimLD21}, we provide a standardized way of reporting that is transferable to other domains.

\cardsName are a tool to acknowledge AI usage openly, similar to how the use of computing infrastructure, external funding, and carbon footprint is acknowledged. 
The community can decide whether using specific models during certain phases of a scientific project (e.g., hypotheses formulation) is reasonable when certain biases exist.
Eventually, analyzing \cardsName across a field of study or area can help to reveal trends in AI usage.

The current version of \cardsName allows for only a limited number of reporters, so we cannot account for many individuals in a scientific project.
In this context, focal points or group leaders should take responsibility for reporting their team's actions. 
The dimensions and properties of \cardsName are a high-level framework for the responsible use and reporting of AI assistance. 
The questions, subcategories, and technical details of the framework should not remain static in the long run.
We invite researchers to build on top of \cardsName and iterate on its concepts.
In the same way, scientific contributions are constantly challenged by the community, AI reporting should be regularly updated to guarantee its relevance and mitigation of outdated practices in the community.
We leave to future work the investigation on how to adapt \cardsName to the specific needs of its users and projects.

\begin{acks}
This work was supported by the Lower Saxony Ministry of Science and Culture and the VW Foundation. 
\end{acks}

\bibliographystyle{ACM-Reference-Format}
\bibliography{sample-base}

\appendix

\section{Ethical considerations}
As AI language models are trained mostly on human-produced data to generate and manipulate textual content, they are not free of the same issues we are subject to (e.g., discrimination and bias).
Thus, using these models might result in inappropriate outcomes that need to be dealt with either during their training phase (e.g., more diverse data) or post-processing (e.g., discarding wrongful text suggestions).
The propagation of unethical content (e.g., racial slurs) can lead to destructive consequences, particularly for vulnerable populations \cite{Spinde2021f}. 

\cardsName cannot guarantee AI will be used to the best of everyone's interest or it will not suggest false claims about sensitive topics and individuals.
However, the human component plays a fundamental role in evaluating altered content and ideas.
It is our responsibility to ``keep in check" the technologies we create and use, so we can take action when necessary.
This does not mean our process is error-free.
Science is a collective work, so our process requires honesty from all parties involved. 
Therefore, if researchers using AI models do not report their actions, our card can propagate bias, unethical statements, and prejudice.

\section{\cardsName for this Paper} \label{appendix:card-this-paper}
Card \ref{card:this-paper} reports the AI Usage of this paper. We used AI for suggestions on the name of the cards, for comparing methods on the theoretical model, and to generate a first version of the abstract, which we did not use in the final manuscript.

\section{\cardsName Template} \label{appendix:cardsample}
Card \ref{card:template} shows a template for \cardsName with details on each subcategory. While these categories describe steps of scientific works, they can be adjusted to other domains as well.

\section{Details on the Questionnaire}
\label{appendix:questionnaire}

\begin{figure}
    \centering
    \includegraphics[width=\columnwidth]{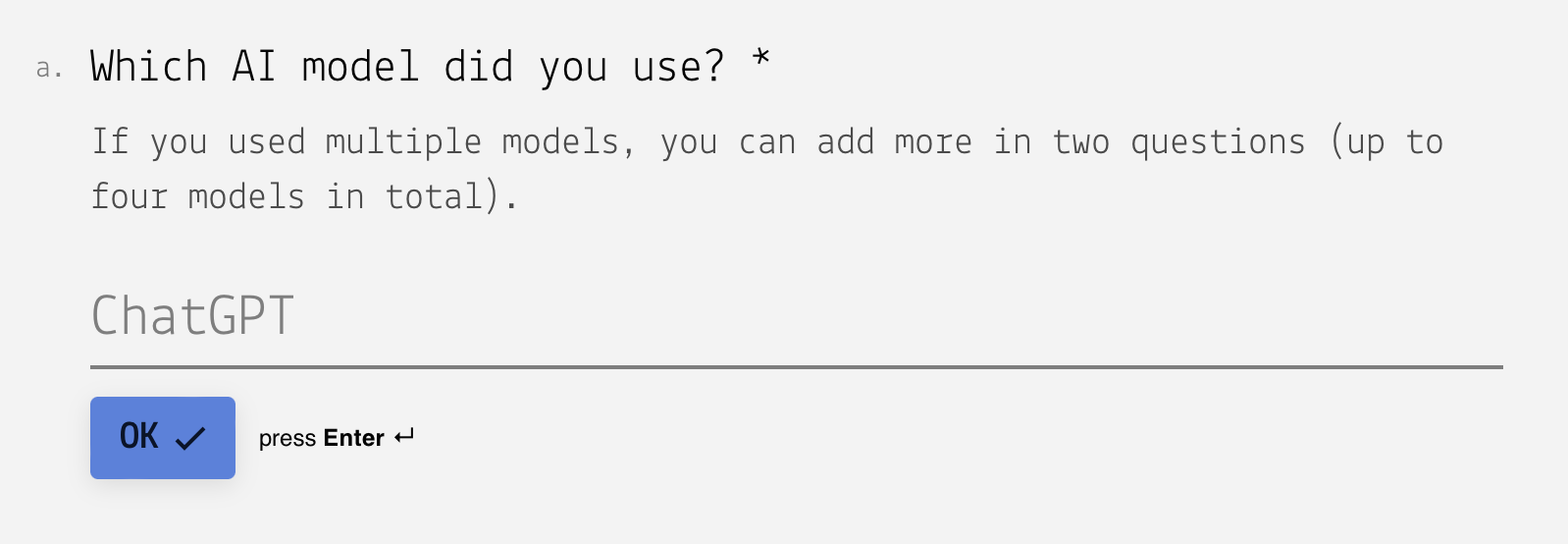}
    \caption{One question of the question set about which AI models were used, when they were used, and in which versions.}
    \label{fig:model_questions}
\end{figure}

\begin{figure}
    \centering
    \includegraphics[width=\columnwidth]{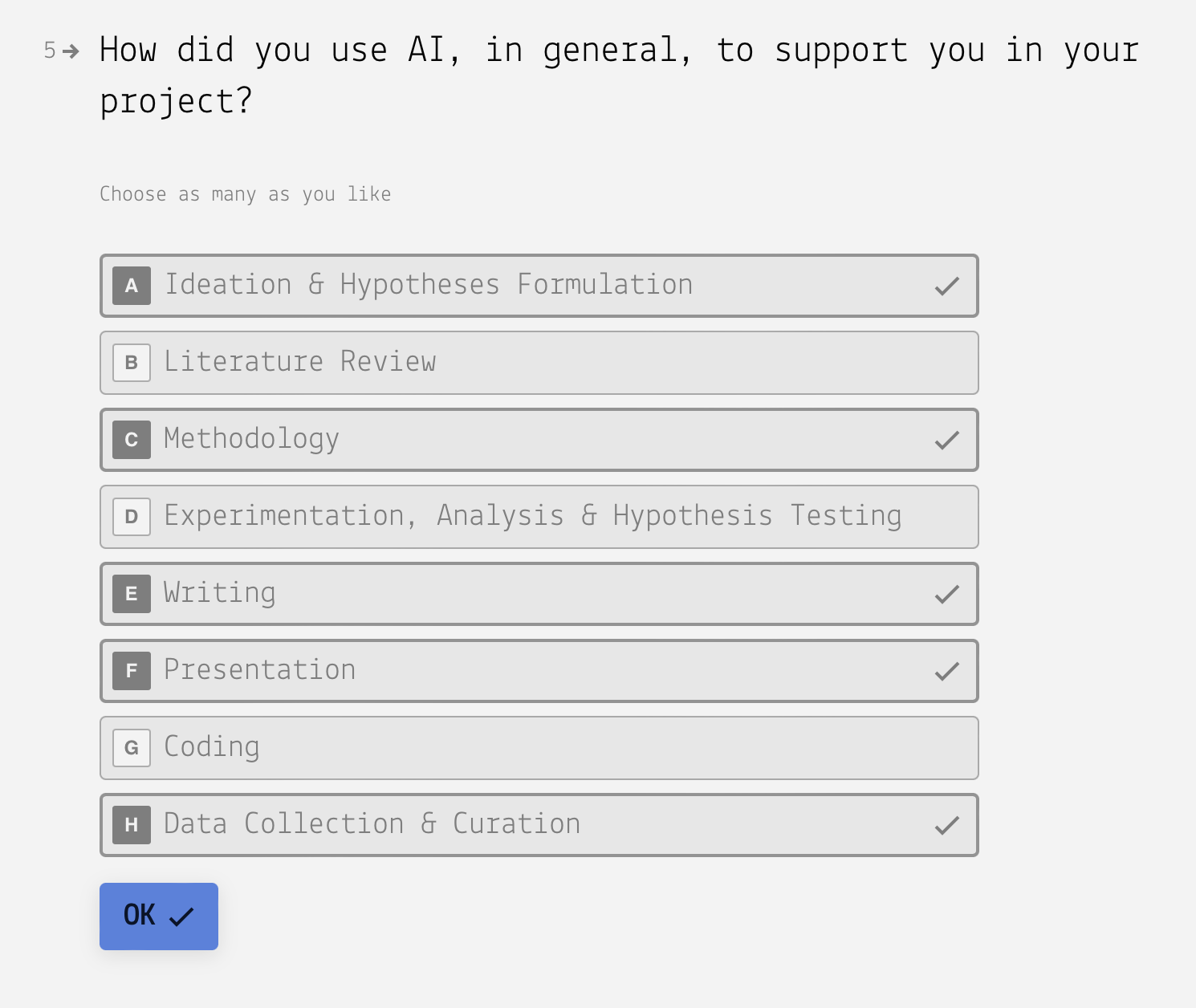}
    \caption{The main categorization in which parts of a work AI was used. This answers the question: ``Where was AI used?''.}
    \label{fig:main_categories}
\end{figure}

\begin{figure}
    \centering
    \includegraphics[width=\columnwidth]{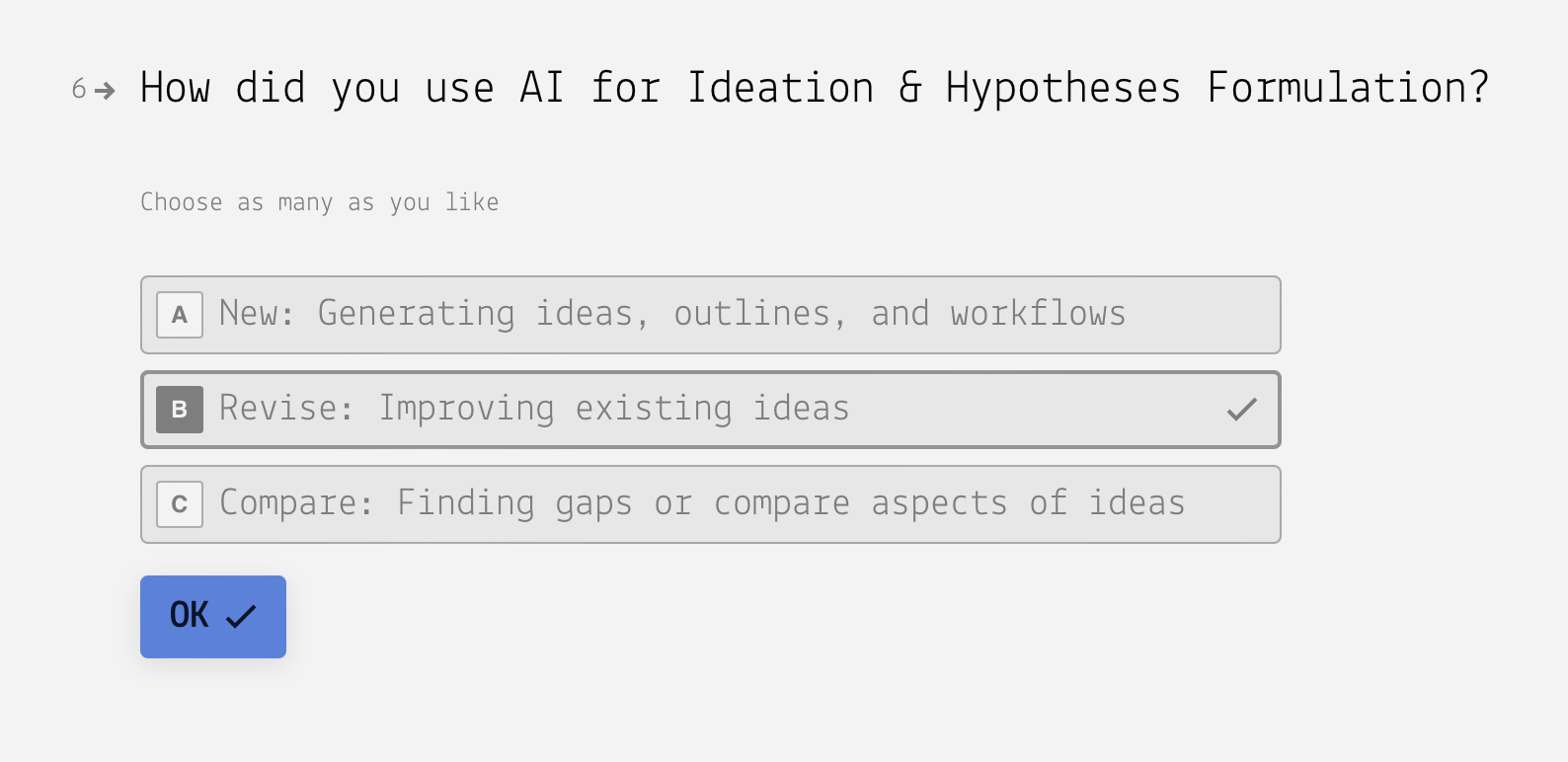}
    \caption{The sub-categories for selected main categories in which AI was used. This answers the question ``How was AI used in this specific aspect of the work?''}
    \label{fig:subcategories}
\end{figure}

\begin{figure}
    \centering
    \includegraphics[width=\columnwidth]{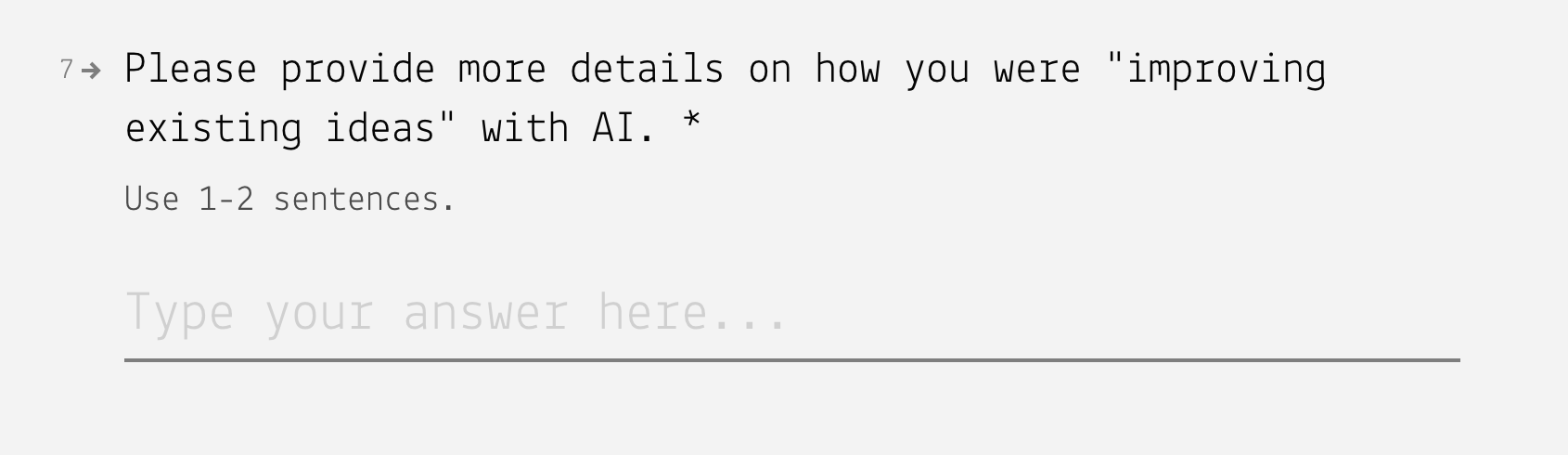}
    \caption{Open text details on each subcategory to mention specifics about how a model was used.}
    \label{fig:subcategories-details}
\end{figure}

\begin{figure}
    \centering
    \includegraphics[width=\columnwidth]{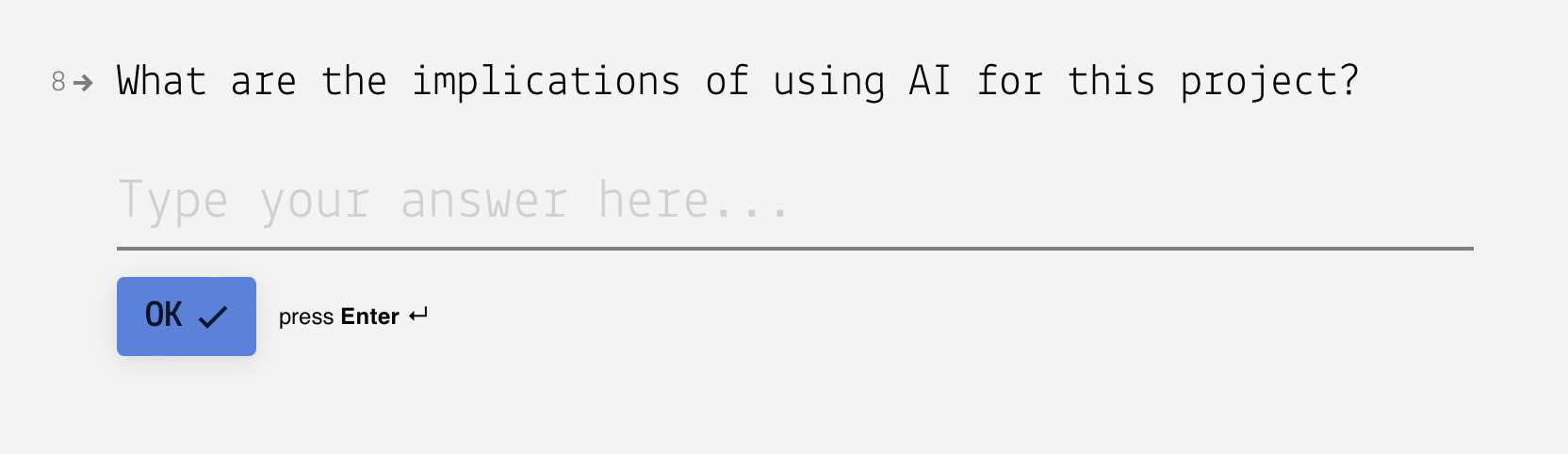}
    \caption{One question of the question set about ethical considerstaions}
    \label{fig:ethics}
\end{figure}

\begin{figure}
    \centering
    \includegraphics[width=\columnwidth]{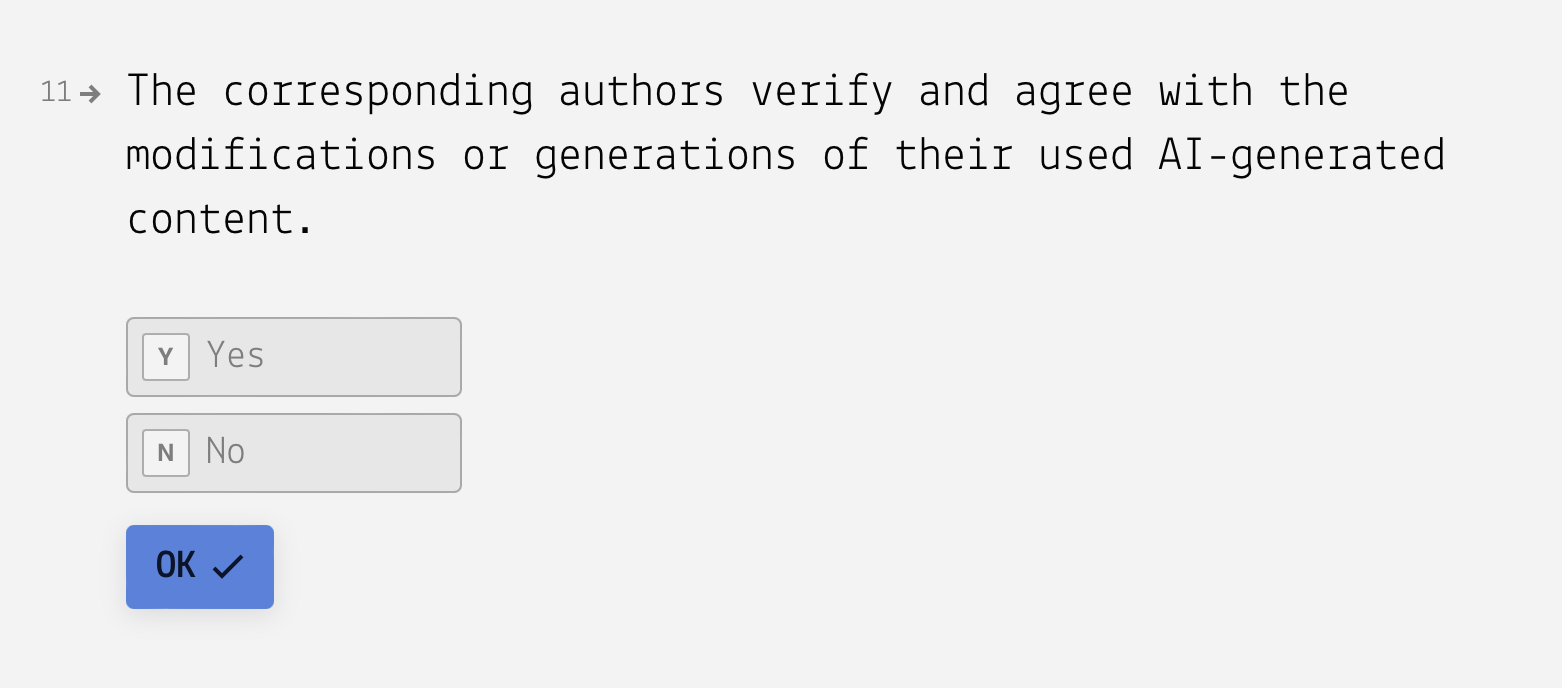}
    \caption{A confirmation that authors approve the used content of AI.}
    \label{fig:integrity}
\end{figure}

\begin{figure}
    \centering
    \includegraphics[width=\columnwidth]{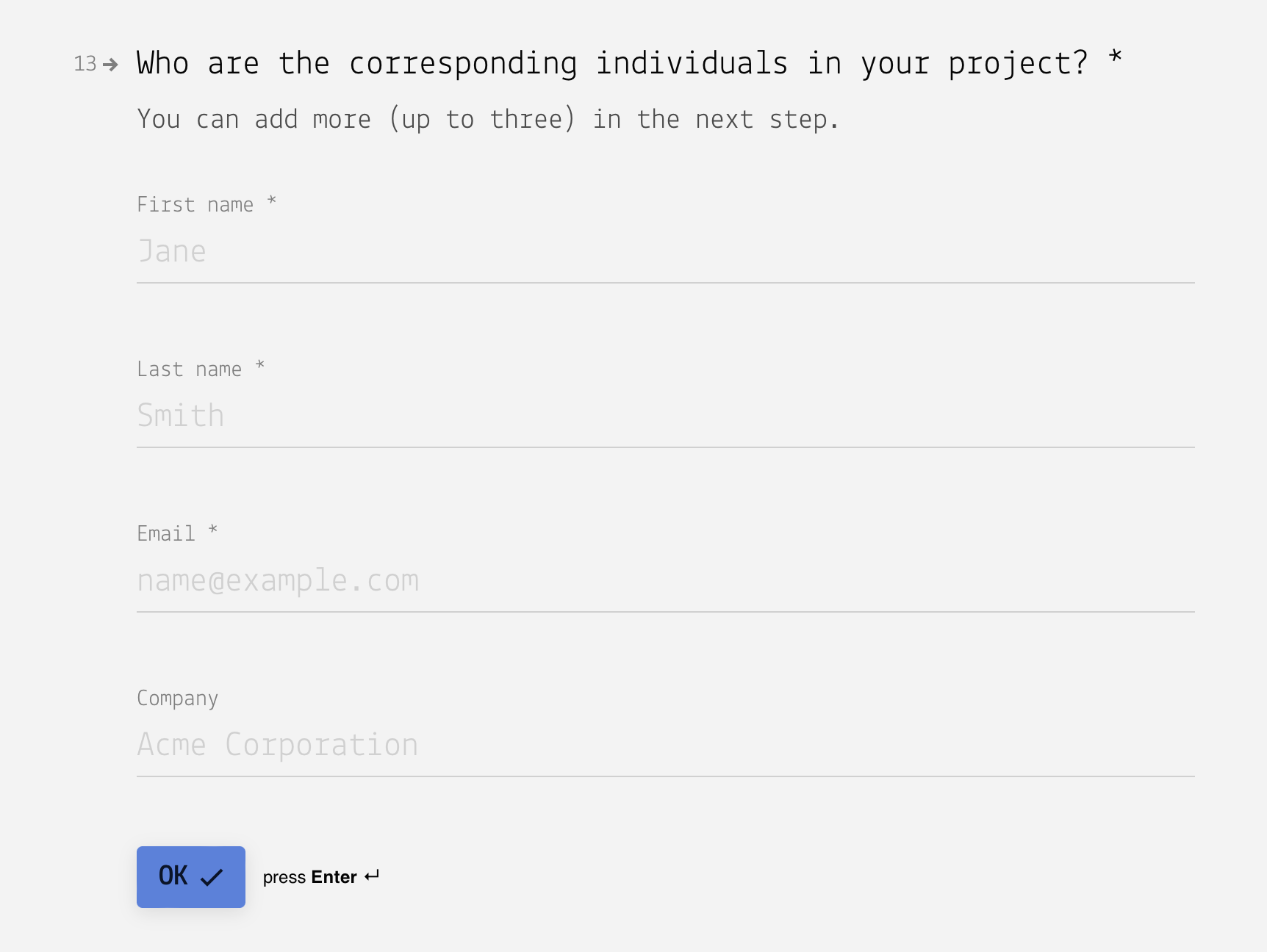}
    \caption{A question of the set for project details to mention the corresponding authors.}
    \label{fig:correspondence}
\end{figure}

\begin{figure}
    \centering
    \includegraphics[width=\columnwidth]{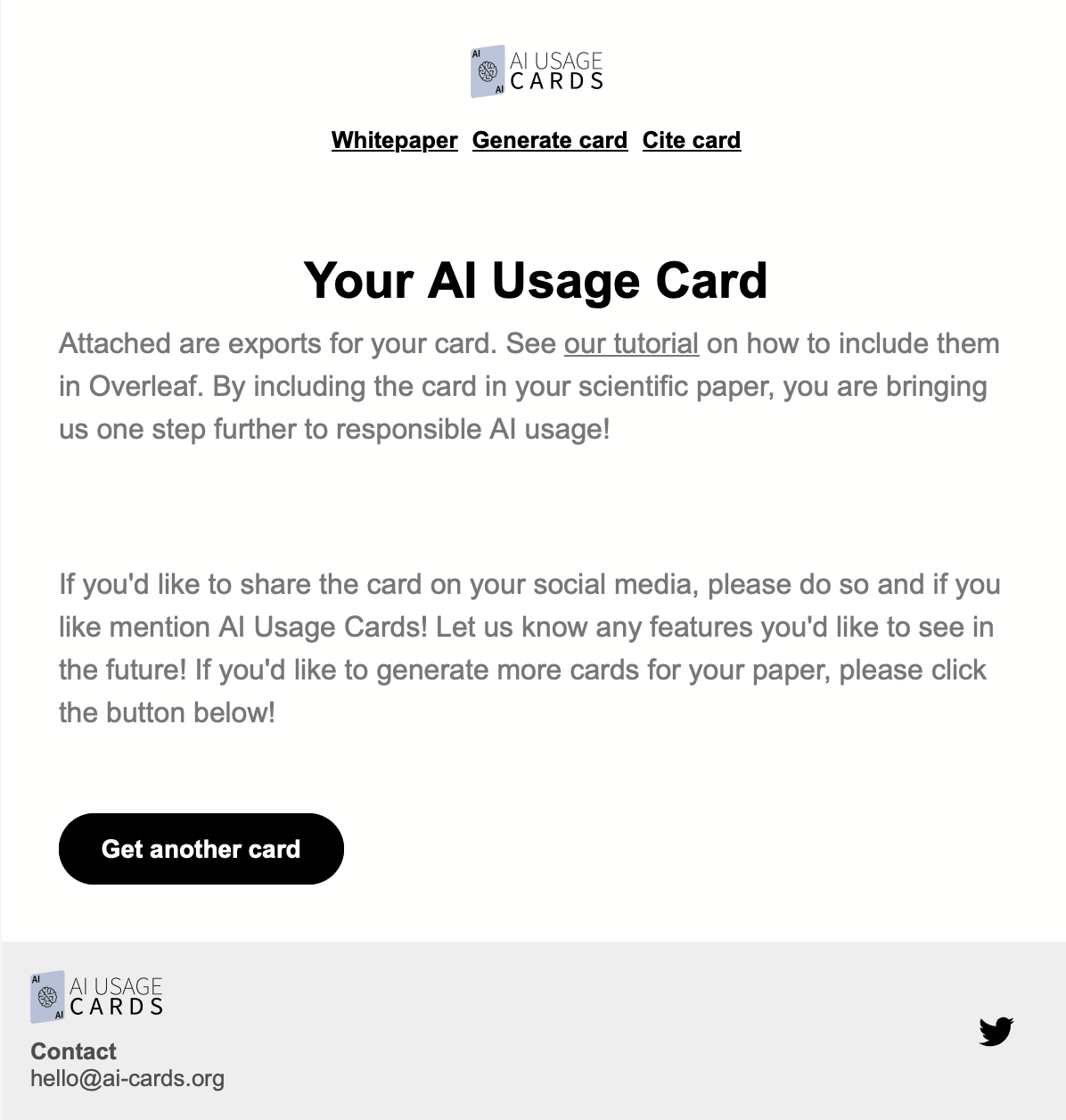}
    \caption{The automatically generated response email to the questionnaire with exports of \cardsName attached and a tutorial on how to include them in scientific works.}
    \label{fig:email}
\end{figure}

We provide a fast process to generate \cardsName by asking users questions.
First, we ask which models have been used in this work by a set of questions, including the model name, date of usage, and model version (see \Cref{fig:model_questions} for an example).
This set of questions can be repeated multiple times. Next, let users choose the main categories in which AI has been used in their work (see \Cref{fig:main_categories}).
Later, we assign each model to a specific category to differentiate whether one model has been used solely during writing while another one was also used for literature review.
The main categories answer the ``where'' of AI usage. Next, based on the selected main categories, we provide unique views to obtain more details on the ``how'' of AI usage.
For example, has AI been used to generate \textbf{new} ideas or to \textbf{revise} or \textbf{compare}? (\Cref{fig:subcategories}).
For each subcategory, users can provide more details in the form of text in the following steps (\Cref{fig:subcategories-details}).
They can explain why it was necessary to use AI, how AI has improved the content, which parts of AI content were used, etc.
Next, users are presented with ethical questions (\Cref{fig:ethics}) and need to approve the used AI-generated content (\Cref{fig:integrity}).
Finally, a set of questions about the project are presented, including the corresponding authors accountable for the AI usage (see \Cref{fig:correspondence} for an example).
After submitting the form, the cards are automatically generated and sent by email with a tutorial on how to include them in scientific works (\Cref{fig:email}).

\onecolumn

\input{assets/card-for-this-project.tex}

\vspace{5mm}

\input{assets/card-template.tex}

\end{document}

%% file: assets/card-for-this-project.tex
{\sffamily
    \centering
    \tcbset{colback=white!10!white}
    \begin{tcolorbox}[
        title={\large \vspace{3mm} \textbf{\textit{\cardsName: Responsibly Reporting AI-generated Content}\vspace{3mm}}},
        breakable,
        boxrule=0.7pt,
        width=.8\paperwidth,
        center,
        skin=bicolor,
        before lower={\footnotesize{AI Usage Card v1.0 \hfill \url{https://ai-cards.org} \hfill \href{https://ai-cards.blind-review.com/whitepaper.pdf}{PDF} | \href{https://ai-cards.blind-review.com/whitepaper.bib}{BibTeX} | \href{https://ai-cards.blind-review.com/whitepaper.xml}{XML} | \href{https://ai-cards.blind-review.com/whitepaper.json}{JSON} | \href{https://ai-cards.blind-review.com/whitepaper.csv}{CSV}}},
        segmentation empty,
        halign lower=center,
        collower=white,
        colbacklower=tcbcolframe]
            
        \footnotesize{
            \begin{longtable}{p{.15\paperwidth} p{.275\paperwidth} p{.275\paperwidth}}
              {\color{LightBlue} \MakeUppercase{Correspondence(s)}} \newline Redacted for anonymity %
              & {\color{LightBlue} \MakeUppercase{Contact(s)}} \newline Redacted for anonymity%
              & {\color{LightBlue} \MakeUppercase{Affiliation(s)}} \newline Redacted for anonymity %
              \\\\
              & {\color{LightBlue} \MakeUppercase{Project Name}} \newline AI Usage Cards for Responsibly Reporting Generated Content 
              & {\color{LightBlue} \MakeUppercase{Key Application(s)}} \newline Artificial Intelligence, Reporting, Responsible AI
              \\\\
              {\color{LightBlue} \MakeUppercase{Model(s)}} \newline ChatGPT
              & {\color{LightBlue} \MakeUppercase{Date(s) Used}} \newline 2023-01-21
              & {\color{LightBlue} \MakeUppercase{Version(s)}} \newline Not used\\\\
              \cmidrule{2-3}\\
      
              {\color{LightBlue} \MakeUppercase{Ideation}} \newline ChatGPT   
              & {\color{LightBlue} \MakeUppercase{Generating ideas, outlines, and workflows}} \newline Not used 
              & {\color{LightBlue} \MakeUppercase{Improving existing ideas}} \newline Gathering more ideas for the name of AI Usage Cards. \\\\
              & {\color{LightBlue} \MakeUppercase{Finding gaps or compare aspects of ideas}} \newline Not used \\\\
              
              {\color{LightBlue} \MakeUppercase{Literature Review}} \newline    
              & {\color{LightBlue} \MakeUppercase{Finding literature}} \newline Not used
              & {\color{LightBlue} \MakeUppercase{Finding examples from known literature}} \newline Not used \\\\
              & {\color{LightBlue} \MakeUppercase{Adding additional literature for existing statements and facts}} \newline Not used
              & {\color{LightBlue} \MakeUppercase{Comparing literature}} \newline Not used \\\\
              \cmidrule{2-3}\\
      
              {\color{LightBlue} \MakeUppercase{Methodology}} \newline ChatGPT   
              & {\color{LightBlue} \MakeUppercase{Proposing new solutions to problems}} \newline Not used
              & {\color{LightBlue} \MakeUppercase{Finding iterative optimizations}} \newline Not used \\\\
              & {\color{LightBlue} \MakeUppercase{Comparing related solutions}} \newline Compare multiple versions of our theoretical model. \\\\
              
              {\color{LightBlue} \MakeUppercase{Experiments}} \newline    
              & {\color{LightBlue} \MakeUppercase{Designing new experiments}} \newline Not used
              & {\color{LightBlue} \MakeUppercase{Editing existing experiments}} \newline Not used \\\\
              & {\color{LightBlue} \MakeUppercase{Finding, comparing, and aggregating results}} \newline Not used \\\\
              \cmidrule{2-3}\\
      
              {\color{LightBlue} \MakeUppercase{Writing}} \newline ChatGPT   
              & {\color{LightBlue} \MakeUppercase{Generating new text based on instructions}} \newline Generated a first version of the abstract which was not used in the final manuscript.
              & {\color{LightBlue} \MakeUppercase{Assisting in improving own content}} \newline Not used \\\\
              & {\color{LightBlue} \MakeUppercase{Paraphrasing related work}} \newline Not used 
              & {\color{LightBlue} \MakeUppercase{Putting other works in perspective}} \newline Not used \\\\
              
              {\color{LightBlue} \MakeUppercase{Presentation}} \newline    
              & {\color{LightBlue} \MakeUppercase{Generating new artifacts}} \newline Not used
              & {\color{LightBlue} \MakeUppercase{Improving the aesthetics of artifacts}} \newline Not used \\\\
              & {\color{LightBlue} \MakeUppercase{Finding relations between own or related artifacts}} \newline Not used \\\\
              \cmidrule{2-3}\\
              {\color{LightBlue} \MakeUppercase{Coding}} \newline    
              & {\color{LightBlue} \MakeUppercase{Generating new code based on descriptions or existing code}} \newline Not used
              & {\color{LightBlue} \MakeUppercase{Refactoring and optimizing existing code}} \newline Not used \\\\
              & {\color{LightBlue} \MakeUppercase{Comparing aspects of existing code}} \newline Not used \\\\
              
              {\color{LightBlue} \MakeUppercase{Data}} \newline    
              & {\color{LightBlue} \MakeUppercase{Suggesting new sources for data collection}} \newline Not used 
              & {\color{LightBlue} \MakeUppercase{Cleaning, normalizing, or standardizing data}} \newline Not used  \\\\
              & {\color{LightBlue} \MakeUppercase{Finding relations between data and collection methods}} \newline Not used  \\\\
              \cmidrule{2-3}\\
      
              {\color{LightBlue} \MakeUppercase{Ethics}} \newline ChatGPT   
              & {\color{LightBlue} \MakeUppercase{What are the implications of using AI for this project?}} \newline Facilitate the AI usage in scientific work (reporting).
              & {\color{LightBlue} \MakeUppercase{What steps are we taking to mitigate errors of AI for this project?}} \newline Careful evaluation of any generated content from the AI model. \\\\
              & {\color{LightBlue} \MakeUppercase{What steps are we taking to minimize the chance of harm or inappropriate use of AI for this project?}} \newline Documentation of suggested content in the scientific document.
              & {\color{LightBlue} \MakeUppercase{The corresponding authors verify and agree with the modifications or generations of their  used AI-generated content}} \newline Yes \\
            \end{longtable}
        }
        \tcblower
    \end{tcolorbox}
    \setcounter{table}{0}
    \captionof{table}{A template for \cardsName.}
    \label{card:this-paper}
}

%% file: assets/card-template.tex
{\sffamily
    \tcbset{colback=white!10!white}
    \begin{tcolorbox}[enhanced,
        title={\large \vspace{3mm} \textbf{AI Usage Card for \textit{Project} - Template \vspace{3mm}}},
        breakable,
        boxrule=0.7pt,
        width=.8\paperwidth,
        center,
        skin=bicolor,
        segmentation empty,
        before lower={\footnotesize{AI Usage Card v1.0 \hfill \url{https://ai-cards.org} \hfill \href{https://ai-cards.org/whitepaper.pdf}{PDF} | \href{https://ai-cards.org/whitepaper.bib}{BibTeX} | \href{https://ai-cards.org/whitepaper.xml}{XML} | \href{https://ai-cards.org/whitepaper.json}{JSON} | \href{https://ai-cards.org/whitepaper.csv}{CSV}}},
        halign lower=center,
        collower=white,
        colbacklower=tcbcolframe]
        
    \footnotesize{
        \begin{longtable}{p{.15\paperwidth} p{.275\paperwidth} p{.275\paperwidth}}
             {\color{LightBlue} \MakeUppercase{Correspondence(s)}} \newline  Author name. 
             & {\color{LightBlue} \MakeUppercase{Contact(s)}} \newline Email address of author. 
             & {\color{LightBlue} \MakeUppercase{Affiliation(s)}} \newline Institution of authors. \\\\
             & {\color{LightBlue} \MakeUppercase{Project Name}} \newline The name of the project. Usually, the paper title. & {\color{LightBlue} \MakeUppercase{Key Application(s)}} \newline The tasks and applications the project. \\\\
             {\color{LightBlue} \MakeUppercase{Model(s)}} \newline Model/Model Card Link \newline Model/Model Card Link
             & {\color{LightBlue} \MakeUppercase{Date(s) Used}} \newline YYYY/MM/DD \newline YYYY/MM/DD 
             & {\color{LightBlue} \MakeUppercase{Version(s)}} \newline Specific version of the model. \newline Specific version of the model. \\\\
             \cmidrule{2-3}\\
             
             {\color{LightBlue} \MakeUppercase{Ideation}} \newline ChatGPT, GPT-3, BERT 
             & {\color{LightBlue} \MakeUppercase{Generating ideas, outlines, and workflows}} \newline When the project direction, topics, outlines, and research questions are generated through prompts or instructions. 
             & {\color{LightBlue} \MakeUppercase{Improving existing ideas}} \newline When existing project ideas, topics, outline, and research questions are either paraphrased, extended, or improved. \\\\
             & {\color{LightBlue} \MakeUppercase{Finding gaps or compare aspects of ideas}} \newline When models are used to identify missing aspects in existing content or compare them. \\\\
             
             {\color{LightBlue} \MakeUppercase{Literature Review}} \newline ChatGPT, GPT-3 
             & {\color{LightBlue} \MakeUppercase{Finding literature}} \newline When unknown related work, supporting literature, or similar is obtained through models. 
             & {\color{LightBlue} \MakeUppercase{Finding examples from known literature}} \newline When examples from a collection of known literature are specified as relevant. \\\\
             & {\color{LightBlue} \MakeUppercase{Adding additional literature for existing statements and facts}} \newline When literature material is suggested to support existing content.
             & {\color{LightBlue} \MakeUppercase{Comparing literature}} \newline When suggested or existing material is compared and analyzed by the model. \\\\
             \cmidrule{2-3}\\
             
             {\color{LightBlue} \MakeUppercase{Methodology}} \newline RoBERTa
             & {\color{LightBlue} \MakeUppercase{Proposing new solutions to problems}} \newline When the method and process for solving the problem are outlined. 
             & {\color{LightBlue} \MakeUppercase{Finding iterative optimizations}} \newline When existing method and process are improved. \\\\
             & {\color{LightBlue} \MakeUppercase{Comparing related solutions}} \newline When existing or generated methods and processes are compared. \\\\
             
             {\color{LightBlue} \MakeUppercase{Experiments}} \newline ChatGPT
             & {\color{LightBlue} \MakeUppercase{Designing new experiments}} \newline When new experiment setups are generated through prompts or instructions. 
             & {\color{LightBlue} \MakeUppercase{Editing existing experiments}} \newline When existing or generated experimental setup is improved. \\\\
             & {\color{LightBlue} \MakeUppercase{Finding, comparing, and aggregating results}} \newline When unseen patterns are suggested using existing or generated results to support analysis. \\\\
             \cmidrule{2-3}\\
             
             {\color{LightBlue} \MakeUppercase{Writing}} \newline GPT-3 
             & {\color{LightBlue} \MakeUppercase{Generating new text based on instructions}} \newline When any text is generated through prompts, questions, or instructions. 
             & {\color{LightBlue} \MakeUppercase{Assisting in improving own content}} \newline When existing text is paraphrased or improved. \\\\
             & {\color{LightBlue} \MakeUppercase{Paraphrasing related work}} \newline When related work content is paraphrased. 
             & {\color{LightBlue} \MakeUppercase{Putting other works in perspective}} \newline When related work is challenged or paraphrased towards a different direction from their original content. \\\\
             
             {\color{LightBlue} \MakeUppercase{Presentation}} \newline DALL E 2, Stable Diffusion 
             & {\color{LightBlue} \MakeUppercase{Generating new artifacts}} \newline When new tables, figures, diagrams, or similar elements are generated through instructions or prompts. 
             & {\color{LightBlue} \MakeUppercase{Improving the aesthetics of artifacts}} \newline When the visual aspects of tables, figures, diagrams, or similar elements are improved. \\\\
             & {\color{LightBlue} \MakeUppercase{Finding relations between own or related artifacts}} \newline When the content of tables, figures, diagrams, or similar elements are compared to uncover unseen relations. \\\\
             \cmidrule{2-3}\\
             
             {\color{LightBlue} \MakeUppercase{Coding}} \newline PaLM
             & {\color{LightBlue} \MakeUppercase{Generating new code based on descriptions or existing code}} \newline When new code is generated based on instructions or prompts. 
             & {\color{LightBlue} \MakeUppercase{Refactoring and optimizing existing code}} \newline When existing or generated code is refactored or its performance optimized. \\\\
             & {\color{LightBlue} \MakeUppercase{Comparing aspects of existing code}} \newline When existing or generated code is compared to uncover unseen patterns or flaws. \\\\
             
             {\color{LightBlue} \MakeUppercase{Data}} \newline T5
             & {\color{LightBlue} \MakeUppercase{Suggesting new sources for data collection}} \newline When datasets, collections, or similar sources are suggested based on instructions or prompts. 
             & {\color{LightBlue} \MakeUppercase{Cleaning, normalizing, or standardizing data}} \newline When any form of noise is removed or mitigated from existing or suggested data. \\\\
             & {\color{LightBlue} \MakeUppercase{Finding relations between data and collection methods}} \newline When models are used to establish any relation between datasets' content and collection methods. \\\\
             \cmidrule{2-3}\\
             
             {\color{LightBlue} \MakeUppercase{Ethics}}
             & {\color{LightBlue} \MakeUppercase{What are the implications of using AI for this project?}} \newline Explain the implications of using AI in the current work scope and its broader impact. 
             & {\color{LightBlue} \MakeUppercase{What steps are we taking to mitigate errors of AI for this project?}} \newline Explain which decisions and actions were taken to minimize or eliminate the use of AI in this project. \\\\
             & {\color{LightBlue} \MakeUppercase{What steps are we taking to minimize the chance of harm or inappropriate use of AI for this project?}} \newline Explain the decisions and actions taken to minimize any form of harm, misuse, and discrimination of the AI model towards any individuals. 
             & {\color{LightBlue} \MakeUppercase{The corresponding authors verify and agree with the modifications or generations of their  used AI-generated content}} \newline Verify that any generated or modified content was approved by the authors involved. This can include facts, statements, ideas, and others.   \\
        \end{longtable}
     }
    \tcblower \raggedright
    \end{tcolorbox}
    \setcounter{table}{1}
    \captionof{table}{A template for \cardsName.}
    \label{card:template}
}